\documentclass[aps,pra,twocolumn,nofootinbib, superscriptaddress,10pt]{revtex4-2}

\usepackage{graphicx}
\usepackage{subfigure}
\usepackage{epstopdf}
\usepackage{amsmath}
\usepackage{amssymb}
\usepackage{amsfonts}
\usepackage{mathrsfs}
\usepackage{theorem}
\usepackage{bm}
\usepackage{url}
\usepackage[T1]{fontenc}
\usepackage{csquotes}
\MakeOuterQuote{"}
\usepackage{multirow}
\usepackage{makecell}
\usepackage{array}
\usepackage{booktabs}

\usepackage{algorithm}
\usepackage{algorithmicx}
\usepackage{algpseudocode}

\usepackage{dcolumn}
\usepackage{color}

\def\vec#1{\mathbf{#1}} 

\newcommand{\Tr}[1]{\mathrm{Tr}\left(#1\right)}
\newcommand{\bn}{\mathbf{n}}
\newcommand{\hbn}{\hat{\mathbf{n}}}
\newcommand{\bz}{\mathbf{z}}
\newcommand{\cH}{\mathcal{H}}
\newcommand{\sym}{\mathrm{sym}}
\newcommand{\dd}{\mathrm{d}}
\newcommand{\abst}{\mathrm{abst}}
\newcommand{\av}{\mathrm{av}}
\newcommand{\id}{1\!\!1}
\newcommand{\hbr}{\hat{\mathbf{r}}}
\newcommand{\ideal}{\mathrm{ideal}}
\newcommand{\ex}{\mathrm{exp}}
\newcommand{\hp}{\hat{p}}
\newcommand{\bQ}{\bar{Q}}
\newcommand{\bPi}{\bar{\Pi}}
\newcommand{\tc}{\tilde{c}}

\newcommand{\BD}{\mathrm{BD}}
\newcommand{\rmi}{\mathrm{i}}
\newcommand{\rme}{\mathrm{e}}

\def\<{\langle}  
\def\>{\rangle}  
\newcommand{\ket}[1]{| #1\>}
\newcommand{\bra}[1]{\< #1|}

\newcommand{\braket}[2]{\langle #1  |#2\rangle}
\newcommand{\ketbra}[2]{|#1\rangle\langle #2|  }

\def\eqref#1{\textup{(\ref{#1})}}  
\newcommand{\eref}[1]{Eq.~\textup{(\ref{#1})}}

\newcommand{\fref}[1]{Fig.~\ref{#1}}

\newcommand{\tref}[1]{Table~\ref{#1}}

\newcommand{\sref}[1]{Sec.~\ref{#1}}

\newcommand{\aref}[1]{Appendix~\ref{#1}}

\newcommand{\bbl}{\bar{\lambda}}

\begin{document}
	\title{Universal device for two-qubit entangled measurements via photonic quantum walks}
	\begin{abstract}
		Sophisticated positive-operator-valued measures (POVMs) are fundamental to obtain a quantum advantage in many informational problems, but difficult to implement.
		We design a universal measurement device composed of cascaded quantum-walk modules able to realize arbitrary entangled POVMs on two qubits. 
		Only parameters in the modules need reprogramming for different POVMs. 
		We apply our device to the task of direction guessing, realizing nine different five-outcome entangled measurements (fidelities above 0.9850) via nine-step photonic quantum walks, recovering optimal direction guessing from non ideal input states.
		Our work marks an important step toward building a universal versatile quantum measurement device able to implement optimal measurements for various quantum information tasks.
	\end{abstract}

	\date{\today}
	
	\author{Wen-Zhe Yan} 
	\affiliation{CAS Key Laboratory of Quantum Information, University of Science and Technology of China, Hefei 230026, People's Republic of China}
	\affiliation{CAS Center For Excellence in Quantum Information and Quantum Physics, University of Science and Technology of China, Hefei 230026, People's Republic of China}
 \affiliation{Hefei National Laboratory, University of Science and Technology of China, Hefei 230088, People's Republic of China}
	\author{Zhibo Hou} 
	\email{houzhibo@ustc.edu.cn}
	\affiliation{CAS Key Laboratory of Quantum Information, University of Science and Technology of China, Hefei 230026, People's Republic of China}
	\affiliation{CAS Center For Excellence in Quantum Information and Quantum Physics, University of Science and Technology of China, Hefei 230026, People's Republic of China} 
 \affiliation{Hefei National Laboratory, University of Science and Technology of China, Hefei 230088, People's Republic of China}
	\author{Jun-Feng Tang} 
	\affiliation{CAS Key Laboratory of Quantum Information, University of Science and Technology of China, Hefei 230026, People's Republic of China}
	\affiliation{CAS Center For Excellence in Quantum Information and Quantum Physics, University of Science and Technology of China, Hefei 230026, People's Republic of China}
  \affiliation{Hefei National Laboratory, University of Science and Technology of China, Hefei 230088, People's Republic of China}
	\author{Guo-Yong Xiang}
	\email{gyxiang@ustc.edu.cn}
	\affiliation{CAS Key Laboratory of Quantum Information, University of Science and Technology of China, Hefei 230026, People's Republic of China}
	\affiliation{CAS Center For Excellence in Quantum Information and Quantum Physics, University of Science and Technology of China, Hefei 230026, People's Republic of China}
  \affiliation{Hefei National Laboratory, University of Science and Technology of China, Hefei 230088, People's Republic of China}
	\author{Chuan-Feng Li}
	\affiliation{CAS Key Laboratory of Quantum Information, University of Science and Technology of China, Hefei 230026, People's Republic of China}
	\affiliation{CAS Center For Excellence in Quantum Information and Quantum Physics, University of Science and Technology of China, Hefei 230026, People's Republic of China}
  \affiliation{Hefei National Laboratory, University of Science and Technology of China, Hefei 230088, People's Republic of China}
	\author{Guang-Can Guo}
	\affiliation{CAS Key Laboratory of Quantum Information, University of Science and Technology of China, Hefei 230026, People's Republic of China}
	\affiliation{CAS Center For Excellence in Quantum Information and Quantum Physics, University of Science and Technology of China, Hefei 230026, People's Republic of China}
 \affiliation{Hefei National Laboratory, University of Science and Technology of China, Hefei 230088, People's Republic of China}	
	\author{Marc-Olivier Renou}
	\email{marc-olivier.renou@icfo.eu}
	\affiliation{ICFO-Institut de Ciencies Fotoniques, The Barcelona Institute of Science and Technology, Castelldefels (Barcelona), Spain}
	\affiliation{Inria Saclay, Bâtiment Alan Turing, 1, rue Honoré d’Estienne d’Orves 91120 Palaiseau}
	\affiliation{CPHT, Ecole polytechnique, Institut Polytechnique de Paris, Route de Saclay – 91128 Palaiseau}
	
	\maketitle

        \section{Introduction}
	
	The manipulation of quantum information outperforms the one of classical information for many informational problems~\cite{Hels76book,NielC00book,Gision02quantum,SelfTestingReview,Renou2018,Bancal2018,Supic2022,GiovLM11}.
	Being able to implement powerful quantum measurements for extracting information is fundamental to obtain such a quantum advantage~\cite{Mass00collective,Zaun11,LiFGK16,ZhuH17U,BagaBGM06S}. 
	
	This requires the ability to realize nonprojective entangled positive-operator-valued measures (POVMs).
	On one hand, entangled (or joint) quantum measurements are represented by measurement operators containing at least one entangled eigenstate~\cite{Renou2018}.
 	A prominent example is collective measurements on identically prepared quantum systems~\cite{PereW91,MassP95}, which view the ensemble as a whole and manipulate entanglement between the copies during the measuring process.
	Entangled measurements offers advantages over separable measurements in information extraction~\cite{PereW91,MassP95,GisiP99}. It is entanglement in the measurements rather than in the states that provides the advantage.
	Entangled measurements have shown significance in numerous tasks such as quantum state estimation~\cite{MassP95,GisiP99,Mass00collective,ZhuH17U}, entanglement purification~\cite{Linden98Pur}, quantum metrology~\cite{VidrDGJ14,RoccGMS17}, overlap estimation~\cite{Fanizza20Bey}, quantum change point detection~\cite{SentBCC16} and quantum thermodynamics~\cite{Perarnau17No,Wu2020Min}. 
	On the other hand,  POVMs~\cite{NielC00book} are the most general description of quantum measurements. A POVM is a set of positive but not necessarily orthogonal operators summing up to the identity. In many scenarios, nonprojective POVMs are more powerful than projective measurements owing to their diverse forms and capabilities~\cite{PereW91,ReneBSC04,Perarnau17No,ZhuH17U,Bergou12Opt,Bagan2012,GendraPRL13,GendraPRA13}.
	A paradigmatic example is a skill called abstention, where the measurements include an inconclusive outcome as in unambiguous discrimination~\cite{IVANOVIC1987257,Mohseni2004optical} and probabilistic quantum metrology~\cite{GendraPRL13}. This is relevant in a context where post-selection is allowed, or if a non-cooperative agent is using apparent losses to increase their estimation score. 
    If a POVM cointains entangled operators, it is an entangled POVM.
	Experimental realization of the powerful entangled POVMs is of paramount importance in practical quantum tasks.
	
	However, how to implement arbitrary entangled POVMs is still an experimental challenge. What is more challenging but desirable is to realize all entangled measurements with one universal measurement device.  
	So far, even for simple two-qubit cases, there only exist some special entangled  measurement devices~\cite{hou2018deterministic,yuan2020direct,Tang2020Exp,Wu2020Min} whose versatility is limited. In particular, there is a recent experimental advance~\cite{conlon2023approaching} that implements arbitrary two-qubit projective entangled measurements on superconducting and trapped-ion systems. Nevertheless, an experimentally amenable method to realize arbitrary nonprojective entangled measurements is still lacking.
	
	In this work, we develop a general recipe for implementing arbitrary two-qubit entangled POVMs using quantum walks~\cite{KurzW13}. 
	We use cascaded quantum-walk modules with programmable coin operators to construct a universal measurement device. Our measurement device is universal in the way that we just reprogram the parameters (wave-plate rotation angles) to realize different entangled measurements.	
	As an experimental application, we focus on the fundamental problem of direction estimation~\cite{MassP95,GisiP99,Bagan2001Optimal,Peres2001,Peres2002,GendraPRA13,Bagan2004,Chiribella2004}. 
	We consider the probabilistic protocol where abstention is used to improve the measurement accuracy.
	Here, the optimal measurement is described by a five-output entangled two-qubit POVM including an abstention operator, and varies with the code state so that cannot be performed on previous special measurement devices. Our universal design makes it possible to implement different optimal measurements on a single device containing a nine-step photonic quantum walk, experimentally recovering the  maximal guessing score even with states which are suboptimal in the standard deterministic protocol~\cite{GisiP99,Bagan2001Optimal,Tang2020Exp}. 
	Note that our design uses two degrees of freedom (DOF) of a single photon to encode two qubits~\cite{Fior04deterministic,Barr08beating,lanyon09sim,Rubi17experimental}. But it can be extended to the collective measurements on two photons by applying the technique of quantum joining~\cite{vite13joining} at the beginning, where the polarization state of two input photons is transferred into a quantum state of a single output photon in the Hilbert space of its polarization and path DOF.

	\section{Realizing arbitrary two-qubit entangled POVMs via quantum walks}\label{sec: QW}
 
	Quantum walks are a powerful tool for implementing general POVMs~\cite{KurzW13,Li2019Implementation}. One-dimensional discrete-time quantum walk is a process in which the evolution of a quantum particle (walker) on a lattice depends on the state of a two-level system (coin). The joint system is characterized by two DOF labeled as $\ket{x,c}$, where $x=...,-1, 0, 1, ...$ denotes the walker position, and $c = 0, 1$ denotes the coin state. The dynamics correspond to several unitary steps. At step $t$, the unitary operator is given by $U(t) = TC(t)$, where  $C(t)=\sum\limits_{x}\ket{x}\bra{x}\otimes C(x,t)$ is determined by site-dependent coin operators $C(x,t)$ representing coin toss, and
	\begin{equation}
		T=\sum\limits_{x}\ket{x+1,0}\bra{x,0}+\ket{x-1,1}\bra{x,1}
	\end{equation}
	is the conditional translation operator, updating the walker position based on the coin state. 
	After $n$ steps, the evolution is given by $U =\prod_{t=1}^{n}TC(t)$. 
	
	It is proven that any POVM on a single qubit, concretely the coin qubit, can be realized by measuring the walker position after certain steps by choosing the coin operators properly \cite{KurzW13}. 
	This has been implemented experimentally \cite{BianLQZ15,ZhaoYKX15}. However, for two-qubit systems, the general method is still lacking.
	
	\begin{figure}[htbp]
		\centering	
		\includegraphics[width=\linewidth]{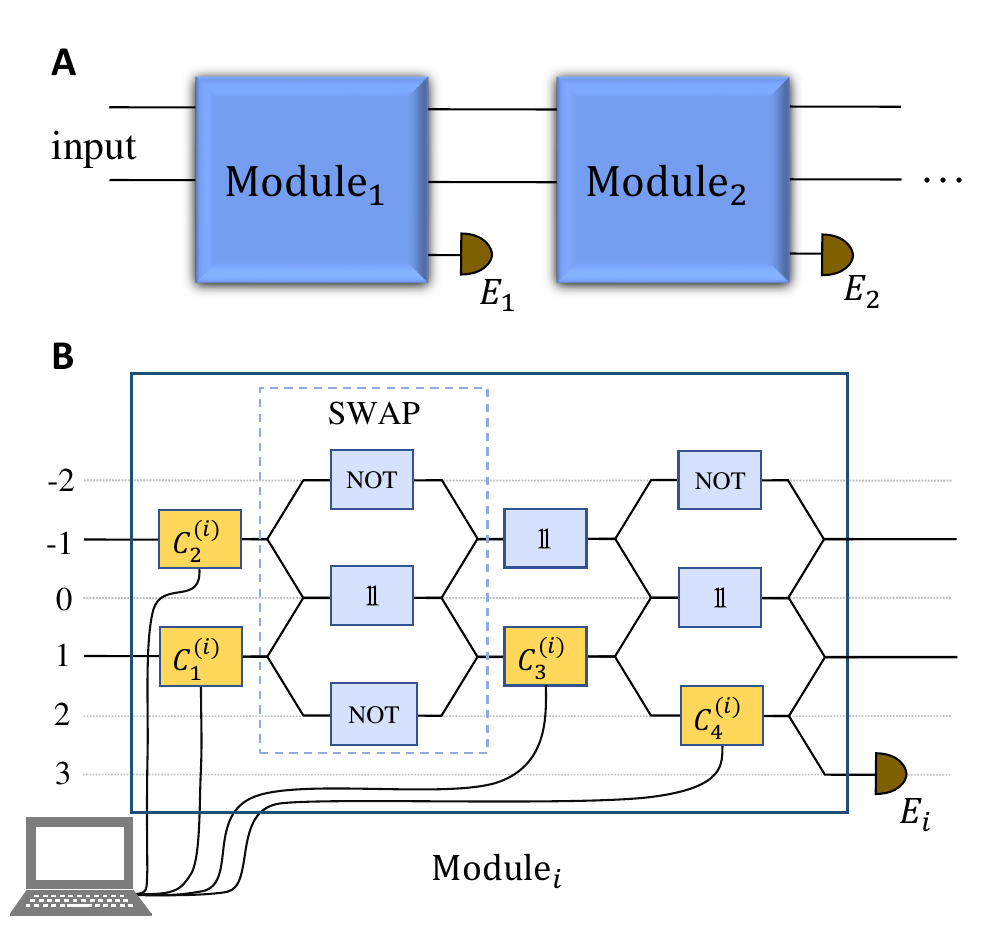}
		\caption{\label{fig: GeneralMethod} General method for implementing an arbitrary two-qubit entangled POVM via cascaded quantum-walk modules. The coin qubit and the walker in positions 1 and -1 are taken as the 2-qubit system of interest. (\textbf{A}) Each POVM element $E_i$ is implemented by a quantum-walk module iteratively. (\textbf{B}) Each module contains a 4-step quantum walk. Positions 3, 2, 0, and -2 act as ancillary levels. Coin operators $C_j^{(i)}(j=1,...,4)$ are programmable and determined by the target POVM. 
		} 
	\end{figure} 
	
	Here, we devise an experimental recipe for implementing arbitrary two-qubit entangled POVMs via quantum walks, as illustrated in \fref{fig: GeneralMethod}. We take the two-level coin qubit and the walker in positions 1 and -1 as the two-qubit system of interest while other positions of the walker act as an ancilla. Consider a rank-1 POVM $\{E_1, E_2,..., E_n\}$ where $E_i=a_i \ketbra{\psi_i}{\psi_i}$, $\ket{\psi_i}$ are normalized two-qubit states and $0< a_i \leq 1$. Each element $E_i$ is implemented by a 4-step quantum-walk module. The fixed coin operators NOT and $\id$ refer to the NOT gate and the identity operator. 
    We adapt the method in \cite{Li2019Implementation} to find the algorithm for calculating the programmable coin operators $C_j^{(i)}(j=1,...,4)$ in the $i$th module, which is elaborated as follows.

        Let $C_4^{(i)}$ be a real matrix in the form of
	\begin{equation}
		C_4^{(i)}=\begin{pmatrix}\alpha_i&\beta_i\\\beta_i&-\alpha_i\end{pmatrix},
	\end{equation}
	where $\alpha_i$ and $\beta_i$ are parameters to be determined.

	Suppose $\ket{\varphi}$ is the two-qubit input state to be measured. Let $\ket{1,0}=\ket{1}$, $\ket{1,1}=\ket{2}$, $\ket{-1,0}=\ket{3}$, $\ket{-1,1}=\ket{4}$. Introduce a matrix $K_i$, which is defined by the relation that the component of the input state entering the $i$th module reads $K_{i}\ket{\varphi}$. Thus $K_1=\id$ and 
	\begin{equation}
		\begin{aligned}					K_{i+1}=&\begin{pmatrix}0&0&1&0\\\beta_i&0&0&0\\0&0&0&1\\0&1&0&0\end{pmatrix}\begin{pmatrix}C_3^{(i)}&0\\0&\id\end{pmatrix}\begin{pmatrix}0&0&1&0\\1&0&0&0\\0&0&0&1\\0&1&0&0\end{pmatrix}\\
		&\begin{pmatrix}C_1^{(i)}&0\\0&C_2^{(i)}\end{pmatrix}K_i.
		\end{aligned}
	\end{equation}  
 
	Introduce a normalized ket $\ket{\eta_i}={b^{-1}_i}K_i^{\dagger+}\ket{\psi_i}$, where $b_i:=\Vert K_i^{\dagger+}\ket{\psi_i}\Vert$, and $K_i^{\dagger+}$ is the Moore-Penrose generalized inverse of $K_i^{\dagger}$.	Then we can determine all the coin operators $C_j^{(i)}$ recursively.
	The coin operators $C_1^{(i)}$, $C_2^{(i)}$ and $C_3^{(i)}$ are chosen such that  	
        \begin{equation}		 \begin{pmatrix}C_3^{(i)}&0\\0&\id\end{pmatrix}\begin{pmatrix}0&0&1&0\\1&0&0&0\\0&0&0&1\\0&1&0&0\end{pmatrix}\begin{pmatrix}C_1^{(i)}&0\\0&C_2^{(i)}\end{pmatrix}\ket{\eta_i}=\ket{1},
        \end{equation} 
    implying that after the third step of the $i$th module, the amplitude entering position 2 is $b^{-1}_i\bra{\psi_i}K_i^{+} K_i\ket{\varphi}$.
	The parameters $\alpha_i$ and $\beta_i$ are given by 
	\begin{equation}
		\alpha_i=b_i\sqrt{a_i}, \quad \beta_i=\sqrt{1-b^2_i a_i}.
	\end{equation} 
 
	Therefore, after the fourth step, the probability of finding the walker at detector ``$E_i$'' reads
	\begin{equation}\label{eq: probability}
		\begin{aligned}
			p_i&=\left|\sqrt{a_i}\bra{\psi_i} K_i^{+} K_i\ket{\varphi}\right|^2=\left|\sqrt{a_i}\bra{\psi_i} P_{K_i} \ket{\varphi}\right|^2\\
			&=a_i\left|\braket{\psi_i}{\varphi}\right|^2=\Tr{E_i\ketbra{\varphi}{\varphi}},
		\end{aligned}
	\end{equation}   
	demonstrating that the desired POVM is realized. Here $P_{K_i}$ is the projector onto supp($K_i$). The third equality follows from the fact $\ket{\psi_i}\in \mathrm{supp}(K_i)$, which can be obtained from the inequality
	\begin{equation}
		E_i=a_i\ketbra{\psi_i}{\psi_i} \leq \id-\sum\limits_{l=1}^{i-1}E_l=K_i^{\dagger} K_i.
	\end{equation}
	Additionally, this inequality also implies that  
	\begin{equation}
		\alpha_i^2=a_i b_i^2=a_i \Vert K_i^{\dagger+} \ketbra{\psi_i}{\psi_i} K_i^{+} \Vert \leq 1,
	\end{equation}
	  thus $C_4^{(i)}$ is well defined. 
	For the last element $E_n$, we just sum the probabilities of finding the walker in positions 1 and -1 at the exit of the $(n-1)$th module. 
 
	It is worth noting that we design a `SWAP' structure that implements a unitary operator 
	\begin{equation}
		\mathrm{U=\begin{pmatrix}0&0&1&0\\1&0&0&0\\0&0&0&1\\0&1&0&0\end{pmatrix}=(\id \otimes NOT) SWAP,}
	\end{equation}
	which makes the originally difficult walker operation easy by swapping the walker qubit to the operation-friendly coin qubit. Additionally, under some specific conditions, the construction in \fref{fig: GeneralMethod} can be further simplified. 
	
	A universal measurement device with reprogramming  can realize a series of measurements without rebuilding the setup. As we will show in the task of direction guessing assisted by abstention, in order to recover the maximal guessing score for different code states, different entangled POVMS are required, which is difficult to realize with previous special measurement devices tailored to a single measurement. 
	
	\begin{figure*}[htbp]
		\centering	
		\includegraphics[width=\textwidth]{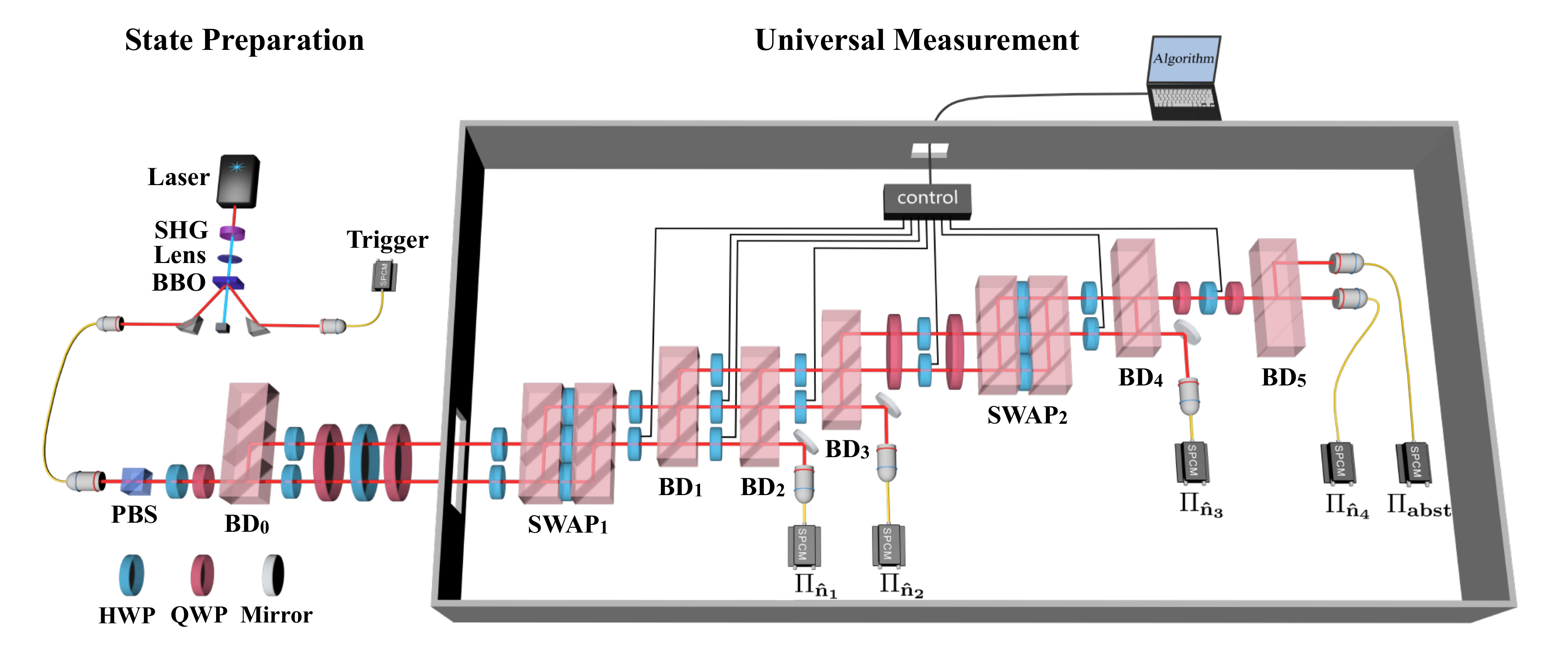}
		\caption{\label{fig: expsetup} Experimental setup for implementing two-qubit entangled measurements and its application in abstention-assisted direction guessing. In the state preparation module, the direction to be estimated is encoded into a two-qubit state in the path and polarization degrees of freedom. Then the universal measurement device performs the optimal POVM via a nine-step quantum walk. The coin operators are realized by a sequence of half-wave plates (HWPs) and quarter-wave plates (QWPs), whose rotation angles are controlled and programmed to implement different measurements. The conditional translation operator $T$ is realized by beam displacers (BDs), each of which includes three polarizing beam splitters (PBSs) coatings separated by 2 mm. The `SWAP' structure contains a compact two-step quantum walk architecture, which is composed of 3 HWPs and 2 BDs. The five  single-photon detectors correspond to the five POVM elements.
		} 
	\end{figure*} 
	
        \section{Optimal direction guessing assisted by abstention}\label{sec: drct-gs}
	
	As an application, we consider the  task of optimal direction guessing assisted by abstention. Concretely, Alice is given a rotated two spin-1/2 state $\ket{\Psi_\bn}:=U_\bn\ket{\Psi_{\bz}}$, where $\ket{\Psi_{\bz}}$ is the initial state, $U_\bn$ is the unitary representation of a rotation mapping states of direction $+\bz$ into states of direction $\bn$, and $\bn$ is selected uniformly at random. 
	She measures it to guess its direction and is allowed not to answer with a probability  $Q=1-\bar{Q}$. The measurement is characterized by a POVM $\Pi=\{\Pi_{\hbn}\}\cup\{\Pi_\abst\}$, where $\Pi_{\hbn}$ generates a guess $\hbn$ and $\Pi_\abst$ corresponds to abstention.	The abstention operator $\Pi_\abst$ is taken rotation invariant, i.e. writes $\Pi_{\abst}=\lambda_0\id_{\cH_0}+\lambda_1\id_{\cH_1}$~\cite{GendraPRL13,GendraPRA13}, where $\cH_{0}$ is the singlet space spanned by $\ket{\psi^-}$, $\cH_{1}$ is the triplet space spanned by $\ket{\psi^+}, \ket{\phi^{\pm}}$, in which $\ket{\phi^\pm}=1/\sqrt{2}(\ket{00}\pm\ket{11}), \ket{\psi^\pm}=1/\sqrt{2}(\ket{01}\pm\ket{10})$, and coefficients $\lambda_0, \lambda_1 \in [0, 1]$ can be optimized to obtain the best guess. We measure Alice's performances with two scores $s\in\{f,\delta\}$. 
	The first is the fidelity score $f(\bn,\hbn)=1/2(1+\bn\cdot\hbn)$.
	The second is the maximum likelihood score characterized by a Dirac distribution $\delta(\bn,\hbn)$~\footnote{The maximum likelihood score should be understood asymptotically. $\delta(\bn,\hbn)=p(\hbn\in\dd\mathcal{A})/|\dd\mathcal{A}|$, where $\dd\mathcal{A}$ is an infinitesimally small subset $\dd\mathcal{A}\subset SU(2)$ containing $\bn$ and $|\dd\mathcal{A}|$ its area.
		For instance, if $SU(2)$ is divided in four parts of equal areas and Alice is exactly able to guess in which of these four parts $\bn$ is, then $\delta(\bn,\hbn)=4$.}, which only measures the likelihood to perform a perfect guess. The average guessing score is computed only on the cases in which she answers, given by \begin{equation}\label{eq: MeanScoreAbst}
		s_\av=\frac{1}{1-Q}\int \dd\bn\dd\hbn s(\bn,\hbn) \Tr{\Pi_{\hbn} \Psi_\bn},
	\end{equation}
	where $\Psi_\bn=\ketbra{\Psi_\bn}{\Psi_\bn}$ and
	$Q=\int \dd\bn\Tr{\Pi_\abst \Psi_\bn}$.
	Write a general initial state as
	\begin{equation}\label{eq:InputState}
		\ket{\Psi_z} = c_0 \ket{\psi^-} + c_1 \ket{\psi^\sym},
	\end{equation}
	where $\ket{\psi^\sym}\in\cH_1$.

	It can be proven that in the standard deterministic protocols where abstention is not allowed, i.e. $\lambda_0=\lambda_1=0$, the optimal fidelity $f_{\max}=(3+\sqrt{3})/6$ is only obtained for $c_0=c_1=1/\sqrt{2}, \ket{\psi^{\sym}}=\ket{\psi^{+}}$, and the optimal maximum likelihood score $\delta_{\max}=4$ is only obtained for $c_0=1/2, c_1=\sqrt{3}/2$ independently of 
	$\ket{\psi^\sym}\in\cH_1$. By contrast, in our probabilistic protocol which succeeds with probability $\bar{Q}$, abstention acts as a filter to discard unfavorable outcomes and makes it possible to recover the optimal guessing score even from suboptimal states (see \aref{app: drct-gs}). Concretely, Alice can always reach $f_{\max}$ at the condition that $c_0,c_1\neq 0, \ket{\psi^{\sym}}=\ket{\psi^{+}}$ and that she is allowed to not answer with probability $Q\geq 1- 2\min(c_0^2,c_1^2)$, by taking $\bbl_k=1-\lambda_k=(\bar{Q})/(2c_k^2)$.
	She can also reach $\delta_{\max}$ at the condition that $c_0,c_1\neq 0$ and that she is allowed to not answer with probability  $Q\geq 1-4\min(c_0^2,c_1^2/3)$, by taking $\bbl_0=\bar{Q}/(4c_0^2),\bbl_1=3\bar{Q}/(4c_1^2)$. 
	The condition $c_0,c_1\neq 0$ is intuitive because the code states should explore the full Hilbert space $\cH=\cH_{0}\oplus\cH_{1}$ to obtain a good encoding property~\cite{GisiP99}.
	In both cases, Alice simulates the behavior of an optimal initial state using abstention.
	
	We now assume that $\ket{\psi^{\sym}}=\ket{\psi^{+}}$. The optimal POVM elements (in addition to $\Pi_{\abst}$) can be taken to be four rank-one operators  $\Pi_{\hbn_i}= \ketbra{\Phi_{\hbn_i}}{\Phi_{\hbn_i}}$, where
	\begin{equation}\label{eq: povm}
		\ket{\Phi_{\hbn_i}}=\frac{\sqrt{\bbl_0}}{2}\ket{\psi^-_{\hbn_i}}+\frac{\sqrt{3\bbl_1}}{2}\ket{\psi^+_{\hbn_i}}, \quad i=1,2,3,4
	\end{equation}
	are sub normalized vectors and $\hbn_i$ points at the four extremities of a tetrahedron. Using the method in \fref{fig: GeneralMethod}, the above two-qubit entangled POVM with any $\bbl_0,\bbl_1 \in [0, 1]$ can be implemented on a single device containing four cascaded quantum-walk modules (16-step quantum walks). Under the restriction $\bbl_1=1$, i.e., abstention is only applied in $\cH_{0}$, the construction can be simplified to a 9-step quantum walk for experimental ease (see \aref{app: simplif}).
	
	\begin{table*}[htbp]
		\centering
		\caption{\label{tab: tomography} \textbf{Overall fidelities of the implemented POVMs.} $\bbl_1=1$ is fixed and $\bbl_0$ is determined by $c_0$ according to the optimization conditions. Each data point is the average over 25 repetitions of 20000 runs. The error bars indicate the standard deviation of 25 repetitions.}
		\begin{tabular}{|p{3cm}<{\centering} | p{3cm}<{\centering} | p{2cm}<{\centering} p{2cm}<{\centering} p{2cm}<{\centering} p{2cm}<{\centering} p{2cm}<{\centering}| }  
			\hline 
			\multirow{3}{*}{likelihood score}
			\rule{0pt}{8.5pt}
			& $c_0$  &  0.5  &  0.6  &  0.7  & 0.8  & 0.9 \\ \cline{2-7}
			\rule{0pt}{8.5pt}
			& $\bbl_0$ & 1 & 0.5926 & 0.3469 & 0.1875 & 0.0782\\ \cline{2-7}
			\rule{0pt}{9pt}
			& overall fidelity &0.9854(1)&0.9857(5)&0.9852(7)&0.9893(7)&0.9850(7)\\ \hline
			\rule{0pt}{8.5pt}
			\multirow{3}{*}{fidelity score}
			\rule{0pt}{8.5pt}
			& $c_0$  &  0.7071  &  0.7571  &  0.8071  & 0.8571  & 0.9071 \\ \cline{2-7}
			\rule{0pt}{8.5pt}
			& $\bbl_0$ & 1 & 0.7446 & 0.5351 & 0.3612 & 0.2153\\ \cline{2-7}
			\rule{0pt}{8.5pt}
			& overall fidelity &0.9854(1)&0.9876(9)&0.9866(8)&0.9856(7)&0.9870(6)\\
			\hline 
		\end{tabular}
	\end{table*}

        \section{Experimental setup}
	
	We experimentally build up the measurement device to implement the entangled POVM of~\eref{eq: povm} and demonstrate the optimal direction guessing with abstention (see \fref{fig: expsetup}). Here, we restrict $\bbl_1=1$ and optimize $\bbl_0$ to recover the maximal score.
	The polarization and path DOF of a single photon act as the coin and walker, respectively. 
	The site-dependent coin operators are realized by a sequence of half-wave plates (HWPs) and quarter-wave plates (QWPs). 
	The conditional translation operator is realized by interferometrically stable beam displacers (BDs), which displace the component with horizontal polarization (H) away from the component with vertical polarization (V).  
	In our experimental setup, we use a structure composed of three bonded polarizing beam splitters (PBSs) as a BD. 
	
	The direction $\vec{n}$ is encoded into a two-qubit state in the path and polarization DOF through a state preparation process (see \aref{app: state-prep}). Next, the universal measurement device performs the optimal POVMs via a nine-step photonic quantum walk. We just reprogram the rotation angles of some wave plates to implement different measurements(see specific angles in \aref{app: Rotation angles}). To reduce the  phase instability of the interferometers, which is the main problem leading to the decrease of precision as the steps of walks increase, we concatenate two BDs and three HWPs respectively set at $45^\circ$, $0^\circ$ and $45^\circ$ as a compact `SWAP' structure. 
	This compactness makes the structure more robust against environmental perturbations. 
	The five single-photon-counting modules (SPCMs) at the end correspond to the five POVM elements. 
	Based on the measurement outcome, Alice gives her answer ${\hbn}$ or abstains.
	
	\section{Experimental results}

	Because we fix $\bbl_1=1$ and optimize $\bbl_0$, The maximal likelihood (resp. fidelity) score can be recovered for $1/2\leq c_0< 1$ (resp. $1/\sqrt{2}\leq c_0< 1$).
	We choose five points of $c_0$ in each situation (see \tref{tab: tomography}), corresponding to a total of nine measurements. 
	
	We first perform quantum measurement tomography to accurately characterize the measurements that were actually realized. 
	We prepare and send 20 input states (the mutually unbiased bases for 2 qubits) to the measurement device. 
	We reconstruct the measurements with the method of~\cite{Fiur01maximum}, obtaining overall fidelities above 0.985 for the implemented POVMs, demonstrating a high-quality realization (see~\tref{tab: tomography}).
	
	\begin{figure}[htbp] 
		\centering	
		\includegraphics[width=\linewidth]{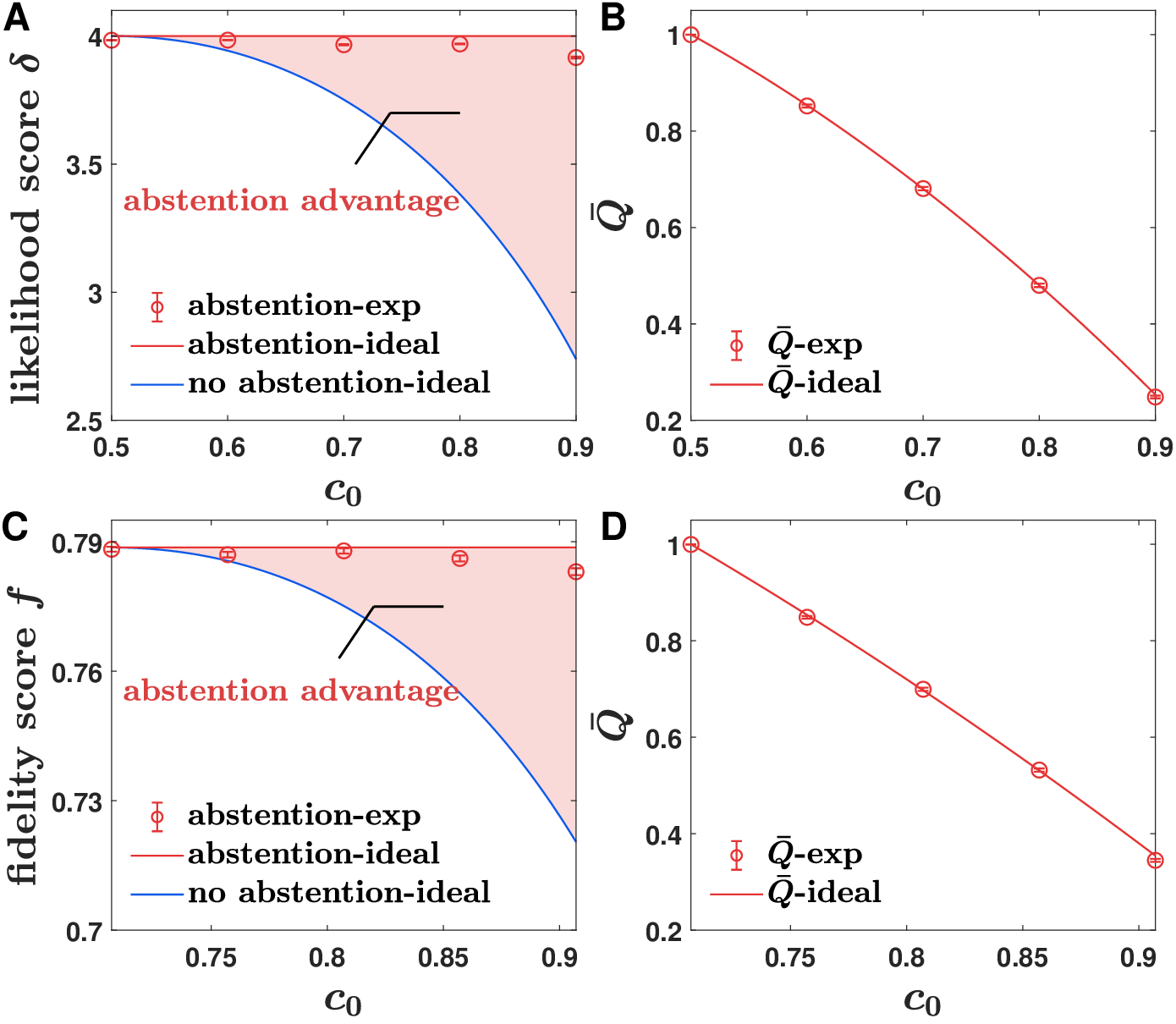}
		\caption{\label{fig: expresults} Experimental results of the optimal direction guessing. For (\textbf{A}) likelihood score and (\textbf{C}) fidelity score, it is experimentally proved that the suboptimal states can also achieve the maximal score with abstention. The corresponding acceptance  rate $\bar{Q}$ in each case is shown in (\textbf{B}) and (\textbf{D}) respectively.
		} 
	\end{figure}

	Next, we implement the optimal direction guessing scheme. 
	The experimental results are illustrated in \fref{fig: expresults} (red circles). 
	The scores we obtain coincide well with theoretical predictions (red solid line). 
	Note that the deviation between the experimental and ideal score is inclined to increase with the growth of $c_0$, due to the term $1/(1-Q)$ in \eref{eq: MeanScoreAbst}, which grows with $c_0$ (see \aref{app: error}).
	We also plot the theoretical scores in no abstention case (blue solid line).
	Hence, for the maximum likelihood score, our experiments almost achieved the maximal $\delta_{\max}=4$ for suboptimal states using abstention, which without abstention would be possible for $c_0=1/2$ only. The abstention advantage increases with the distance between the suboptimal states and the optimal state. Specifically, for the suboptimal state corresponding to $c_0=0.9$, we obtain an experimental score of 3.9162, to be compared to the theoretical value of 2.7390 without abstention: we reduce the deviation from $\delta_{\max}$ by a factor of about 15.
	Similarly, for the fidelity score, we almost recover the maximal $f_{\max}=(3+\sqrt{3})/6\approx{0.7887}$ for suboptimal states using abstention.
	For the suboptimal state corresponding to $c_0=0.9071$, we obtain an experimental score of 0.7831, to be compared to the theoretical value of 0.7204 without abstention: we reduce the deviation from $f_{\max}$ by a factor of about 12. 
	The corresponding acceptance rates $\bar{Q}$ are shown in \fref{fig: expresults}(B,D).
	
	\section{Summary}
	
	We proposed a universal method for realizing arbitrary two-qubit POVMs using quantum walks. We experimentally implemented nine optimal measurements in the direction guessing task on a single device with high fidelities, demonstrating the malleability of our method for concretely implementing complex many-output POVMs. 

	The application is focused on optimal orienting with a two-qubit input state.
	Here, a nonzero component of the state in the singlet space allows for better encoding of the direction in the full Hilbert space. 
	A new effect is theoretically predicted for three-qubit input states. 
	In that case, the Hilbert space decomposes into a spin $\frac{1}{2}$ with multiplicity two and a spin $\frac{3}{2}$.
	One can in principle exploit the multiplicity of the spin $\frac{1}{2}$ to perform a dense coding encryption of the rotation in the input state, which allows for better direction guessing scores~\cite{Chiribella2004,Bagan2004}.
	Being able to witness this non intuitive effect (it was first overlooked in the literature~\cite{Chiribella2004}, see e.g.~\cite{Peres2001,Bagan2001Optimal,Peres2002}) is an interesting experimental challenge, which would require to manipulate and measure three-qubit states. By taking more paths of the walker or using two-dimensional quantum walks, our method can be further extended to realizing entangled POVMs on multi-qubit states. Integrated optics\cite{wang20integrated} is a potential platform for more sophisticated many-path quantum walks.

    \acknowledgements
	We thank Nicolas Gisin, Antonio Acín and Enky Oudot for valuable discussions. The work at the University of Science and Technology of China is supported by the National Natural Science Foundation of China (Grants Nos. 12134014, 61905234, 11974335, and 11774334), the Key Research Program of Frontier Sciences, CAS (Grant No. QYZDYSSW-SLH003), the Innovation Program for Quantum Science and Technology (Grant No. 2021ZD0301203), USTC Research Funds of the Double First-Class Initiative (Grant No. YD2030002007) and the Fundamental Research Funds for the Central Universities (Grant No. WK2470000035).
	M.-O.R. is supported by the grant PCI2021-122022-2B financed by MCIN/AEI/10.13039/501100011033 and by the European Union NextGenerationEU/PRTR and the Swiss National Fund Early Mobility Grants P2GEP2\_19144/2, and acknowledges the Government of Spain (FIS2020-TRANQI and Severo Ochoa CEX2019-000910-S [MCIN/ AEI/10.13039/501100011033]), Fundació Cellex, Fundació Mir-Puig, Generalitat de Catalunya (CERCA, AGAUR SGR 1381) and the ERC AdG CERQUTE.

    \appendix
    \section{Optimal states and measurements in direction guessing tasks}\label{app: drct-gs}

    In this section, we provide details for computing the maximal score and the optimal states and measurements in the direction guessing tasks mentioned in \sref{sec: drct-gs}. 
    
    \subsection{Standard deterministic direction guessing}
    
    The rotation operator can be written as $U_\bn=u_\bn \otimes u_\bn$, where $u_\bn\in SU(2)$ takes the direction $\vec{+z}$ into $\vec{n}$. The guessing score $s\in\{f,\delta\}$ is group invariant, that is $s(\bn,\hbn)=s(u_\bn,u_{\hbn})=s(u_{\hbn}^{-1}u_\bn,\id)\equiv s(u_{\hbn}^{-1}u_\bn)$. One can restrict Alice's POVM measurement to be covariant, that is $\Pi_{\hbn}=U_{\hbn}\Pi_z U_{\hbn}^\dagger$~\cite{Chiribella2011}. The mean score is given by  
	\begin{equation}\label{eq:MeanScore}
		\begin{aligned}
			s_\av&=\int \dd\bn\dd\hbn s(\bn,\hbn) \Tr{\Pi_{\hbn} \Psi_\bn}\\
			&=\int \dd\bn\dd\hbn s(u_{\hbn}^{-1}u_\bn) \Tr{U_{\hbn}\Pi_z U_{\hbn}^\dagger\cdot U_\bn\Psi_z U_\bn^\dagger}\\
			&=\int \dd \bn  s(\bn, +\bz) \Tr{\Pi_{z}U_\bn\Psi_z U_\bn^\dagger}\\
			&=\Tr{\Pi_z F},
		\end{aligned}
	\end{equation}
    where $F=\int \dd \bn s(\bn, +\bz) U_\bn\Psi_z U_\bn^\dagger$. In the third equality we substitute $u_{\hbn}^{-1}u_\bn \rightarrow u_\bn$.
	
    Let $\Pi_z=\ketbra{\eta}{\eta}$. The normalization condition $\int \dd \hbn U_{\hbn}\Pi_z U_{\hbn}^\dagger=\id$ gives
	\begin{equation}\label{eq: seed}
		\ket{\eta}= \ket{\psi^-}+\sqrt{3} \ket{\phi^{\sym}},
	\end{equation}
    where only $\ket{\phi^{\sym}}\in\cH_1$ can be optimized. Note that as we integrate over some probability density, $\ket{\eta}$ is unnormalized.
	
    \subsubsection{maximum likelihood score}
	
    With the maximum likelihood score $\delta(\bn,\hbn)$, we have 
	\begin{equation}
		F_\delta=\int \dd \bn \delta(\bn, +\bz) U_\bn\Psi_z U_\bn^\dagger=\Psi_z.
	\end{equation}
	Writing $\delta_\av$ the average score, we have
	\begin{equation}
            \begin{aligned}
                \delta_\av&=\Tr{\Pi_zF_\delta}=\Tr{\Pi_z\Psi_z}=|\braket{\eta}{\Psi_z}|^2\\
                &=|c_0+\sqrt{3}c_1\braket{\phi^{\sym}}{\psi^{\sym}}|^2.
            \end{aligned}
	\end{equation}
    Hence, for a given initial state $\ket{\Psi_z}$ described by \eref{eq:InputState}, the corresponding optimal measurement is given by imposing $\ket{\phi^{\sym}}= \ket{\psi^{\sym}}$ in $\ket{\eta}$. 
	Then, we have
	\begin{equation}
		\delta_\av=|c_0+\sqrt{3}c_1|^2.
	\end{equation}
    Hence, optimal states are for $c_0=1/2, c_1=\sqrt{3}/2$, with which we obtain the optimal maximum likelihood score
	\begin{equation}
		\delta_{max}=4.
	\end{equation}
	
    \subsubsection{fidelity score}	
    For general case $\ket{\psi^{\sym}}=\alpha\ket{\phi^+}+\beta\ket{\psi^+}+\gamma\ket{\phi^-}$, we have
	\begin{equation}\label{eq: Fforfidelity}
            \begin{aligned}	
		    F_f(\alpha,\beta,\gamma)=&\frac{|c_0|^2}{2}\id_{\cH_0}+\frac{|c_1|^2}{6}\id_{\cH_1}+\frac{1}{6}(c_0 c_1^* \beta^*\ketbra{\psi^-}{\psi^+} \\
            &+ \frac{|c_1|^2\alpha \gamma^*}{2}(\ketbra{\phi^+}{\phi^-}+\ketbra{\phi^-}{\phi^+})+ h.c.).
            \end{aligned}
	\end{equation}
    Let $\ket{\phi^{\sym}}=a\ket{\phi^+}+b\ket{\psi^+}+c\ket{\phi^-}$. The average score is given by
	\begin{equation}
		\begin{aligned}	
			 f_\av(\alpha,\beta,\gamma)=&\Tr{\Pi_zF_f(\alpha,\beta,\gamma)}\\=&\frac{1}{2}+\frac{\sqrt{3}}{6}(c_0 c_1^* \beta^*b+\sqrt{3} |c_1|^2\alpha \gamma^* Re(a^*c)+ c.c.).	
		\end{aligned}
	\end{equation}	
    To maximize $f_\av$, we can restrict ourselves to real positive coefficients (due to the triangular inequality). Then
	\begin{equation}
		\begin{aligned}
			f_\av&=\frac{1}{2}+\frac{\sqrt{3}}{3}(c_0 c_1 \beta b+ \sqrt{3} c_1^2 \alpha \gamma ac)\\
			&\leq \frac{1}{2}+\frac{\sqrt{3}}{3}(c_0 c_1 \beta b+ \frac{\sqrt{3}}{4}c_1^2(1-\beta^2)(1-b^2)),
		\end{aligned}
	\end{equation}
    where we used $\alpha \gamma\leq \frac{\alpha^2+\gamma^2}{2}=\frac{1-\beta^2}{2}$, and a similar bound over $ac$.
    The inequality is saturated  if $\alpha=\gamma=\sqrt{\frac{1-\beta^2}{2}}$ and $a=c=\sqrt{\frac{1-b^2}{2}}$.
    We parameterize $c_0$ and $c_1$ as $c_0=\cos\theta,c_1=\sin\theta$, where $\theta \in [0,\frac{\pi}{2}]$. To optimize $f_\av$, we need to find the maximum of the ternary function
        \begin{equation}
		\begin{aligned}
			f_\av(\beta,b,\theta)=&\frac{1}{2}+\frac{\sqrt{3}}{3}(\cos\theta \sin\theta \beta b\\&+\frac{\sqrt{3}}{4} \sin^2 \theta (1-\beta^2)(1-b^2)).
		\end{aligned}
	\end{equation}
	The only stationary points is:
	\begin{align*}
		\frac{\partial{f_\av}}{\partial{\beta}}=\frac{\partial{f_\av}}{\partial{b}}=\frac{\partial{f_\av}}{\partial{\theta}}=0 \quad \Rightarrow \theta=\frac{\pi}{3},  b=\frac{1}{\sqrt{3}}, \beta=\frac{1}{\sqrt{3}}.
	\end{align*}
    However, $f_\av(\beta=\frac{1}{\sqrt{3}},b=\frac{1}{\sqrt{3}},\theta=\frac{\pi}{3})=\frac{2}{3}$  is not maximal. Indeed, the global maximum is obtained on the boundary (the surface) of the domain  $[0,1]\times[0,1]\times[0,\frac{\pi}{2}]$, for $\beta=1,b=1,\theta=\frac{\pi}{4}$, with the optimal fidelity score being 
	$f_{\max}=f_\av(\beta=1,b=1,\theta=\frac{\pi}{4})=\frac{3+\sqrt{3}}{6}$.
	
    Hence the optimal state is given by $\ket{\Psi_z}=\frac{\ket{\psi^-}+\ket{\psi^+}}{\sqrt{2}}=\ket{01}$, corresponding to two anti-aligned spins, and the optimal measurement is given by imposing $\ket{\phi^{\sym}} = \ket{\psi^{+}}$ in $\ket{\eta}$.
	
    \subsubsection{measurements with finite number of outputs}
	
    Here we consider the problem of finding an optimal measurement which only requires a finite number of outcomes. Suppose we have a \emph{finite} POVM $\{O_{\hbr}\}$ given by $O_{\hbr}=c_{\hbr} U_{\hbr}\ketbra{\mu}{\mu}U_{\hbr}^\dagger$ 
    where $\ket{\mu}$ is a normalized state and $c_{\hbr}$ are positive numbers. Then the average score is given by 
	\begin{equation}
		\begin{aligned}
			s_\av&=\sum_{\hbr}\int_\bn\dd \bn s(\bn,\hbr) \Tr{O_{\hbr} \Psi_\bn} \\
			&=\sum_{\hbr} \Tr{\int_\bn \dd \bn s(u_{\hbr}^{-1}u_\bn) c_{\hbr}U_{\hbr}\ketbra{\mu}{\mu}U_{\hbr}^\dagger U_\bn \Psi_z U_\bn^\dagger}\\
			&=\sum_{\hbr} \Tr{c_{\hbr}\ketbra{\mu}{\mu} \int_\bn \dd \bn s(\bn,+\bz)U_\bn \Psi_z U_\bn^\dagger }\\
			&=\sum_{\hbr} c_{\hbr}\Tr{\ketbra{\mu}{\mu}  F}\\
			&=4\Tr{\ketbra{\mu}{\mu} F}.
		\end{aligned}
	\end{equation}
    The last  equality follows from the completeness condition $\sum_{\hbr}O_{\hbr}=\id$ which gives $\sum_{\hbr} c_{\hbr}=4$ by tracing both sides. Taking $\ket{\mu}= \frac{1}{2}\ket{\psi^-}+\frac{\sqrt{3}}{2} \ket{\phi^{\sym}}$,  we reach the same result as \eref{eq:MeanScore}. Thus the optimization  of $\ket{\phi^{\sym}}$ and the initial state takes the same results as the covariant POVM case. We just need to find a set of $\{U_{\hbr}\}$ and $\{c_{\hbr}\}$ to generate a \emph{finite} POVM from $\ketbra{\mu}{\mu}$.
	
    In the case of $\ket{\phi^{\sym}}=\ket{\psi^+}$, i.e., $\ket{\mu}= \frac{1}{2}\ket{\psi^-}+\frac{\sqrt{3}}{2} \ket{\phi^+}$, the POVM can be taken to be the Elegant Joint Measurement $\Pi_{\hbn_i}^0= \ketbra{\Phi_{\hbn_i}^0}{\Phi_{\hbn_i}^0}$, where
	\begin{equation}\label{eq: povm0}
		\ket{\Phi_{\hbn_i}^0}=\frac{1}{2}\ket{\psi^-_{\hbn_i}}+\frac{\sqrt{3}}{2}\ket{\psi^+_{\hbn_i}}, \quad i=1,2,3,4
	\end{equation}
    and $\hbn_i$ points at the four extremity of a tetrahedron. 

    \subsection{Direction guessing assisted by abstention}
    
    Let $\bPi_\abst=\id-\Pi_\abst=\bbl_0\id_{\cH_0}+\bbl_1\id_{\cH_1}$, where $\bbl_j=1-\lambda_j$. Introducing the new POVM elements
	\begin{equation}\label{eq: newpovm}
		\tilde{\Pi}_{\hbn}=\bPi_\abst^{-1/2}\Pi_{\hbn}\bPi_\abst^{-1/2},
	\end{equation}
    and the new family of states
	\begin{equation}
		\ket{\tilde\Psi_\bn}=\left(\frac{\bPi_\abst}{\bQ}\right)^{1/2}\ket{\Psi_\bn},
	\end{equation}
    the average score given by \eref{eq: MeanScoreAbst} can be rewritten as
	\begin{equation}
		\begin{aligned}
			s_\av&=\int \dd\bn\dd\hbn s(\bn,\hbn) \Tr{\tilde{\Pi}_{\hbn} \tilde\Psi_\bn}\\
			&=\int \dd\bn\dd\hbn s(u_{\hbn}^{-1}u_\bn) \Tr{U_{\hbn}\tilde{\Pi}_z U_{\hbn}^\dagger\cdot U_\bn\tilde\Psi_z U_\bn^\dagger}\\
			&=\int \dd \bn  s(\bn,+\bz) \Tr{\tilde{\Pi}_z U_\bn\tilde\Psi_z U_\bn^\dagger},
		\end{aligned}
	\end{equation}
    implying that we are back to the standard estimation problem with POVM seed $\tilde{\Pi}_{z}$ and initial state $\ket{\tilde\Psi_z}=\tc_0\ket{\psi^-}+\tc_1\ket{\psi^{\sym}}$, where 
	\begin{equation}
		\tc_i=\sqrt{\frac{\bbl_i }{\bQ}}c_i.
 	\end{equation}

    \subsubsection{maximum likelihood score}
	
	Similar to the standard estimation problem, it is best to take $\tilde{\Pi}_{z}=\ketbra{\eta}{\eta}$ with $\ket{\eta}= \ket{\psi^-}+\sqrt{3} \ket{\psi^{\sym}}$, and the optimization problem becomes:
	\begin{equation}
		\delta_{max}=\max_{\stackrel{|\tc_j|\leq |c_j|/\sqrt{\bQ}}{|\tc_0|^2+|\tc_1|^2=1}} |\tc_0+\sqrt{3}\tc_1|^2.
	\end{equation}
	Hence, Alice cannot improve the maximal score $\delta_{max}=4$ whatever the initial state and the abstention rate. However, with $\bQ\leq \min(4c_0^2,\frac{4}{3}c_1^2)$, Alice can always reach $\delta_{\max}$ when she tunes $\bbl_0,\bbl_1$ such that $\tc_0=\frac{1}{2}, \tc_1=\frac{\sqrt 3}{2}$. This is obtained for 
	\begin{equation}\label{eq: optimal}
		\bbl_0=\frac{\bQ}{4 c_0^2}, \bbl_1=\frac{3\bQ}{4 c_1^2}.
	\end{equation}
	Hence, Alice can reach $\delta_{\max}$ if she is allowed to not answer sufficiently often, at the unique condition that $c_0, c_1 \neq 0$.
	
    \subsubsection{fidelity score}
	
	It is best to take  $\ket{\tilde\Psi_z}=\frac{1}{\sqrt{2}}\ket{\psi^-}+\frac{1}{\sqrt{2}}\ket{\psi^{+}}$ and $\tilde{\Pi}_{z}=\ketbra{\eta}{\eta}$ with $\ket{\eta}= \ket{\psi^-}+\sqrt{3} \ket{\psi^{+}}$.
    With $\bQ\leq \min(2c_0^2,2c_1^2)$, Alice can always reach $f_{max}=\frac{3+\sqrt{3}}{6}$ when she tunes $\bbl_0,\bbl_1$ such that $\tc_0=\tc_1=\frac{1}{\sqrt{2}}$. This is obtained for 
	\begin{equation}\label{eq: optimal2}
		\bbl_i=\frac{\bQ}{2 c_i^2}, \quad i=1,2.
	\end{equation}
	at the unique condition that $c_0, c_1 \neq 0$.
 
    \subsubsection{optimal measurements with finite number of outputs}
	
	Using \eref{eq: povm0} and \eref{eq: newpovm},
	one can show that in the case of $\ket{\phi^{\sym}}=\ket{\psi^+}$, the optimal POVM elements (in addition to $\Pi_{\abst}$) can be taken to be four rank one elements $\Pi_{\hbn_i}= \ketbra{\Phi_{\hbn_i}}{\Phi_{\hbn_i}}$, where
	\begin{equation}\label{eq: povm-app}
		\ket{\Phi_{\hbn_i}}=\frac{\sqrt{\bbl_0}}{2}\ket{\psi^-_{\hbn_i}}+\frac{\sqrt{3\bbl_1}}{2}\ket{\psi^+_{\hbn_i}}, \quad i=1,2,3,4.
	\end{equation}

    \section{Simplification of the measurement device for the optimal abstention POVM}\label{app: simplif}
     
\begin{figure}[htbp]
            \center{\includegraphics[width=\linewidth]{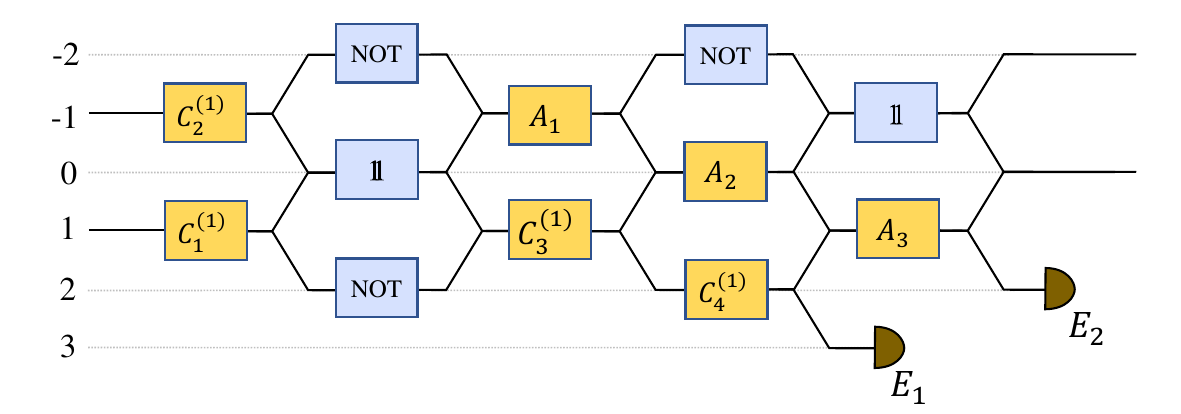}}
            \caption{\label{fig: simplify} Simplification of the combination of $\mathrm{Module_1}$ and $\mathrm{Module_2}$ of the measurement device under the condition of \eref{eq: condition}.}
        \end{figure}

    In this section, we explain how we simplify our universal measurement device to realize the optimal abstention POVM given in \eref{eq: povm} (also in \eref{eq: povm-app}) with $\bbl_1=1$. \fref{fig: GeneralMethod} shows a quantum walk construction for implementing an  arbitrary two-qubit POVM $\{E_1, E_2,..., E_n\}$ where $E_i=a_i \ket{\psi_i}\bra{\psi_i}$. According to the algorithm presented in ~\sref{sec: QW}, the first three steps in $\mathrm{Module_1}$ can be seen as a projection on $\ket{\psi_1}$ (because $K_1=\id$ and $b_1=1$), where the component of $\ket{\psi_1}$ is completely tranformed into $\ket{1}$ and then enters path 2 after in third step.  Utilizing this projection, we decompose $\ket{\psi_2}$ as $\ket{\psi_2}=p\ket{\psi_1}+q\ket{\psi_1^\perp}$. Then the combination of $\mathrm{Module_1}$ and $\mathrm{Module_2}$ can be simplified into the structure shown in  \fref{fig: simplify} under the condition
        \begin{equation}\label{eq: condition}
            a_2|q|^2+\frac{a_2|p|^2}{1-a_1}=1.
        \end{equation}  

    Here, coin operators $C_1^{(1)}$, $C_2^{(1)}$ $C_3^{(1)}$and $C_4^{(1)}$ are still determined by the algorithm in ~\sref{sec: QW}.
    $A_1$ and $A_2$ are choosen such that 
        \begin{equation}
            \begin{aligned}           &\begin{pmatrix}1&0\\0&0\\0&0\\0&1\end{pmatrix}A_2\begin{pmatrix}0&0&1&0\\0&1&0&0\end{pmatrix}\begin{pmatrix}C_3^{(1)}&0\\0&A_1\end{pmatrix}\\
            &\begin{pmatrix}0&0&1&0\\1&0&0&0\\0&0&0&1\\0&1&0&0\end{pmatrix}\begin{pmatrix}C_1^{(1)}&0\\0&C_2^{(1)}\end{pmatrix}\ket{\psi_1^\perp}=\begin{pmatrix}1\\0\\0\\0\end{pmatrix},
            \end{aligned}
        \end{equation} 
    implying that the component of $\ket{\psi_1^\perp}$ is tranformed into $\ket{1}$ after the fourth step. 
    The coin operator $A_3$ is set as 
        \begin{equation}
            A_3=\begin{pmatrix}\sqrt{a_2}q&\frac{\sqrt{a_2}p}{\sqrt{1-a_1}}\\\frac{\sqrt{a_2}p}{\sqrt{1-a_1}}&-\sqrt{a_2}q\end{pmatrix}.
        \end{equation} 
    Thus the amplitude entering path 2 after the fifth step reads
        \begin{equation}
            \begin{aligned}
                &(\sqrt{a_2}q\bra{\psi^\perp}+\frac{\sqrt{a_2}p}{\sqrt{1-a_1}}\sqrt{1-a_1}\bra{\psi_1})\ket{\varphi}\\=&\sqrt{a_2}(q\bra{\psi_1^\perp}+p\bra{\psi_1})\ket{\varphi}\\=&\sqrt{a_2}\braket{\psi_2}{\varphi}, 
            \end{aligned}
        \end{equation}  
    demonstrating the implementation of $E_2$.
    
\begin{figure}[htbp]
		\center{\includegraphics[width=\linewidth]{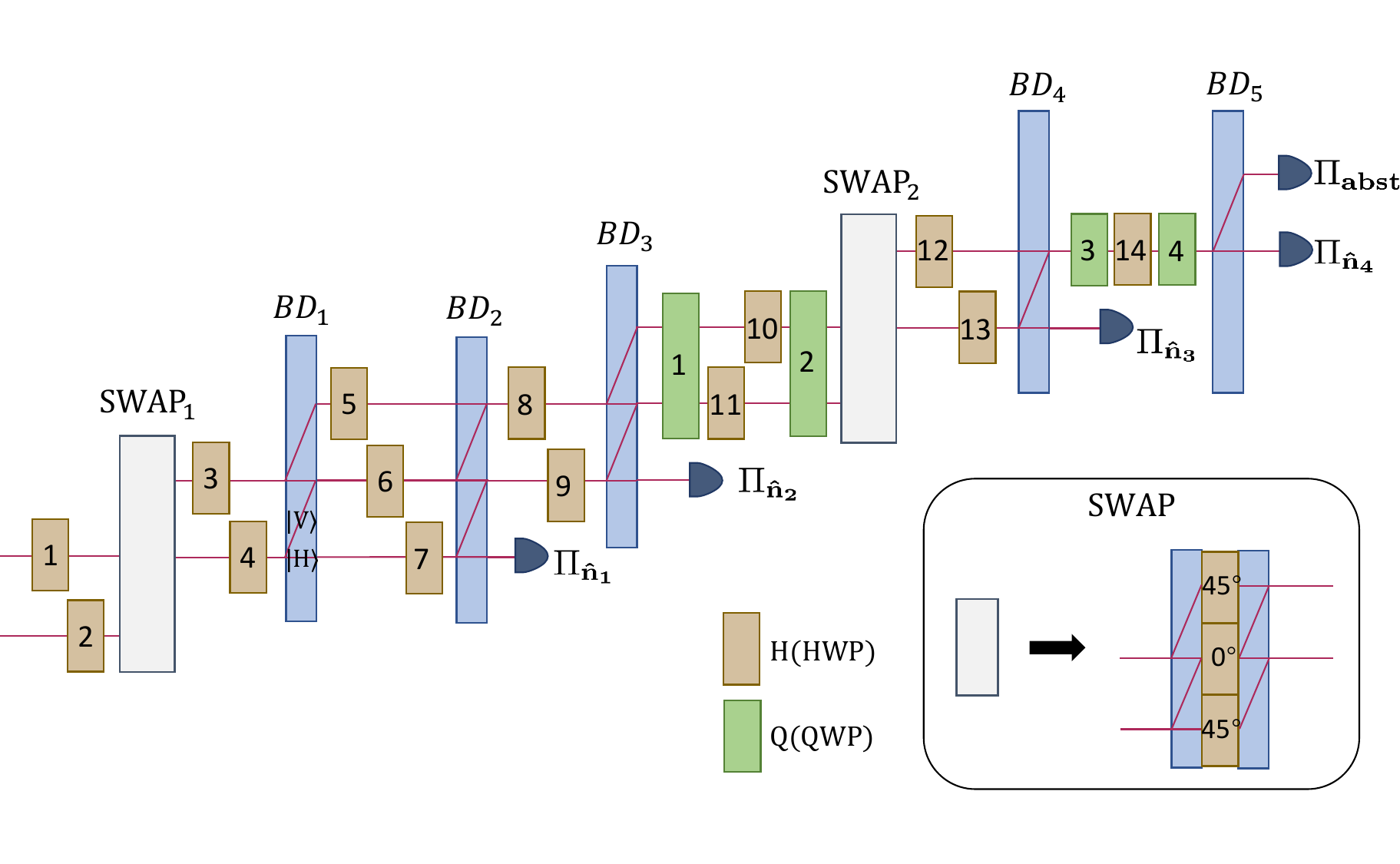}}
		\caption{\label{fig: walk} Realization of the 2-qubit abstention POVMs with $\bbl_1=1$ using 9-step photonic quantum walks.}
	\end{figure}

    Considering the optimal abstention POVM $\{\Pi_{\hbn}\}\cup\{\Pi_\abst\}$, if $\bbl_1=1$, the quantum walk construction can be simplified because the condition of \eref{eq: condition} is satisfied. Additionally, the fourth step in $\mathrm{Module_3}$ and $\mathrm{Module_4}$ can be removed because $C_4^{(3)}=C_4^{(4)}=\id$. Moreover, noting that $C_1^{(4)}=C_2^{(4)}=\mathrm{NOT}$, the first two step in $\mathrm{Module_4}$ can be removed by replacing the coin operator $\id$ in the third step of $\mathrm{Module_3}$ with $\mathrm{NOT}$. Finally, the optimal abstention POVM with $\bbl_1=1$ can be realized with a nine-step quantum walk, shown in \fref{fig: expsetup} ( also in \fref{fig: walk}).

    \section{Experimental state preparation}\label{app: state-prep}
    
    In this section, we give experimental details about how we prepare the encoding state using the state preparation module shown in \fref{fig: expsetup}.
    A light pulse from a mode-locked Ti-sapphire laser at 780 nm passes through a frequency doubler, and then pumps a 2-mm-long BBO crystal. 
    After the type-II phase-matched spontaneous parametric down conversion (SPDC) process, a pair of photons at 780 nm is created. 
    The two photons are coupled to two single-mode fibers respectively. One photon acts as a trigger, while the other is used as a single-photon source and initialized in polarization H by a PBS. 
    After passing a HWP and a QWP, any polarization state can be generated. 
    $\BD_0$ translates the polarization state into the path DOF, thus the walker qubit is prepared. 
    The following combinations of the 5 wave  plates transform Bloch vectors  $\{\vec{+z},\vec{-z}\}$ of the polarization state at positions 1 and -1 into desired $\{\vec{m_1},\vec{m_2}\}$ to prepare the coin qubit.

    Note that for the maximum likelihood score, only the ability to perform perfect guesses matters. 
    Hence, we only consider sending each of the four tetrahedral directions $\{\hbn_1, \hbn_2, \hbn_3, \hbn_4\}$ to the measurement device~\footnote{A rigorous justification of our experimental procedure is the following. 
		At each trial, Alice chooses a redefinition of direction $\bz$ and sends direction $\{\hbn_1, \hbn_2, \hbn_3, \hbn_4\}$ to the referee in charge of computing $\delta(\bn,\hbn)$. In case $\bn\notin\{\hbn_1, \hbn_2, \hbn_3, \hbn_4\}$, the referee attributes score 0 to Alice without even sending the state $\ket{\psi_\bn}$ to here. 
		If not, he sends $\ket{\psi_\bn}$ and check if Alice made the correct guess. This procedure can be understood asymptotically, given a discretisation of the sphere in small subsets $\dd\mathcal{A}$}. 
    And for the fidelity score, we restrict the input states to be one of the six directions $\{\vec{\pm x}, \vec{\pm y}, \vec{\pm z}\}$, which is an alternative to the direction randomly selected from the unit sphere. This way to select the input direction of the state corresponds to the same operator $F_f(\ket{\psi^{\sym}}=\ket{\psi^+})$:
	\begin{equation}
		\begin{aligned}
			&F_f(\ket{\psi^{\sym}}=\ket{\psi^+})=\sum_{\bn \in \{\vec{\pm x}, \vec{\pm y}, \vec{\pm z}\}}\frac{1}{6} \cdot f(\bn) U_\bn\Psi_z U_\bn^\dagger\\
			=&\frac{|c_0|^2}{2}\id_{\cH_0}+\frac{|c_1|^2}{6}\id_{\cH_1}+\frac{c_0c_1^*}{6}(\ketbra{\psi^-}{\psi^+}+ c.c.),
		\end{aligned}
	\end{equation}
    giving the same result as \eref{eq: Fforfidelity}. 

    \section{Rotation angles of the wave plates in the measurement device}\label{app: Rotation angles}
    In this section, we present the specific rotation angles of the wave plates in the measurement module in \fref{fig: expsetup} (also in \fref{fig: walk}).
    In our experiment, we choose the tetrahedral directions as
	\begin{equation}
		\begin{aligned}
			\ket{\hbn_1}=&\ket{0},\qquad \ket{\hbn_2}=\frac{\rmi}{\sqrt{3}}(\ket{0}+\sqrt{2}\ket{1}), \\ \ket{\hbn_3}=&\frac{\rmi}{\sqrt{3}}(\ket{0}+\rme^{\frac{2\pi}{3}\rmi}\sqrt{2}\ket{1}),\\ \ket{\hbn_4}=&\frac{\rmi}{\sqrt{3}}(-\ket{0}+\rme^{\frac{\pi}{3}\rmi}\sqrt{2}\ket{1}).
		\end{aligned}
	\end{equation}
    Because we let $\bbl_1=1$, for the maximum likelihood score, the optimization condition [\eref{eq: optimal}]  is achievable when $\frac{1}{2}\leq c_0 < 1$, and for the fidelity score, the optimization condition [\eref{eq: optimal2}] is achievable when $\frac{\sqrt{2}}{2}\leq c_0 < 1$. The coefficient $c_0$ of the input state we choose in the experiment and the corresponding rotation angles of the wave plates are specified in \tref{tab: fixed angles} to \ref{tab: angles in fidelity}. Here some rotation angles are constant whatever $\bbl_0$ is taken, and some are dependent on $\bbl_0$.
	
	\begin{table}[htbp]
		\centering
		\caption{\label{tab: fixed angles} Rotation angles ($^\circ$)of the wave plates in \fref{fig: walk} which are constant.}
		\begin{tabular*}{1\linewidth}{@{\extracolsep{\fill}} c c c c c c c c c c c}
			\hline
			\rule{0pt}{8.5pt}
			H1  &  H2  &H3  &  H5  &  H8  &  H10  &  H12  &  Q1  &  Q2  &  Q3  &  Q4  \\
			\hline
			\rule{0pt}{8.5pt}
			0&45&67.5&45&0&45&135&0&0&0&0\\
			\hline
		\end{tabular*}
	\end{table}
	
	\begin{table*}[htbp]
		\centering
		\caption{\label{tab: angles in maxlikelihood} Specific $c_0$ of the input state we choose for the maximum likelihood score and the corresponding optimized $\bbl_0$ and the rotation angles ($^\circ$) of the wave plates dependent on $\bbl_0$.}   
		\begin{tabular*}{1\textwidth}{@{\extracolsep{\fill}} c c c c c c c c c}
			\hline
			\rule{0pt}{8.5pt}
			$c_0$  &  $\bbl_0$  &  H4  &  H6  &  H7  &  H9  &  H11  &  H13  &  H14  \\
			\hline
			\rule{0pt}{8.5pt}
			0.5 & 1 & 37.5 & -17.6322  &  0 &  0 &  90 & 67.5 & 135    \\
			0.6 & 0.5926 & 34.4812 & -14.9358  &  9.3055 &  -9.8396 &  76.5853 & 63.7945 & 154.832    \\
			0.7 & 0.3469 & 31.8908 & -12.2404 &  11.9161 & -13.107 &  72.5750 & 60.2494  & 161.9564  \\
			0.8 & 0.1875 & 29.5181 &  -9.4659  & 13.3941 & -15.1616  & 70.2017 &  56.7066 & 167.1705   \\
			0.9 & 0.0782 & 27.0854 &  -6.3509 &  14.3445 & -16.5887 &  68.6211  & 52.8111 & 81.8811 \\
			\hline
		\end{tabular*}
	\end{table*}
	
	\begin{table*}[htbp]
		\centering
		\caption{\label{tab: angles in fidelity} Specific $c_0$ of the input state we choose for the fidelity score and the corresponding optimized $\bbl_0$ and the rotation angles ($^\circ$) of the wave plates dependent on $\bbl_0$.}  
		\begin{tabular*}{1\textwidth}{@{\extracolsep{\fill}} c c c c c c c c c}
			\hline
			\rule{0pt}{8.5pt}
			$c_0$  &  $\bbl_0$  &  H4  &  H6  &  H7  &  H9  &  H11  &  H13  &  H14  \\
			\hline
			\rule{0pt}{8.5pt}
			0.7071 & 1 & 37.5 & -17.6322  &  0 &  0 &  90 & 67.5 & 135    \\
			0.7571 &0.7446&	35.7409&	-16.1181&	7.3188&	-7.5701&	79.5303&	65.3951&	150.1793\\
			0.8071 &0.5351&	33.9481&	-14.4101&	9.9664&	-10.6312&	75.5878&	63.0929&	156.4935\\
			0.8571 &0.3612&	32.0684&	-12.4365&	11.7771&	-12.9225&	72.7939&	60.5034&	161.5285\\
			0.9071 & 0.2153&29.9985&	-10.0501&	13.1451&	-14.8025&	70.6083&	57.4457&	166.1771\\
			\hline
		\end{tabular*}
	\end{table*}

    \section{Error analysis of the experimental score}\label{app: error}
	
	In \fref{fig: expresults}, the experimental deviation of the average score $|\Delta s|=|s_{\mathrm{exp}}-s_{\mathrm{ideal}}|$ increases with the growth of the abstention rate $Q$. We show here that this comes from the amplification of the experimental error by the term $\frac{1}{1-Q}$ in \eref{eq: MeanScoreAbst}. 
	
    As mentioned in \aref{app: state-prep}, for the maximum likelihood score we input each of the four directions $\{\hbn_1, \hbn_2, \hbn_3, \hbn_4\}$ to the measurement device, and for the fidelity score, we input each of the six directions $\{\mathbf{\pm x}, \mathbf{\pm y}, \mathbf{\pm z}\}$. 
    Therefore, the ideal score is given by
	\begin{equation}\label{eq: score_ideal}
		s_\ideal=\frac{1}{1-Q_\ideal}\frac{1}{k}\sum\limits_{j=1}^{k}\sum\limits_{i=1}^{4} s(\bn_j,\hbn_i) \Tr{\Pi_{\hbn_i} \Psi_{\bn_j}},
	\end{equation}
	where $Q_\ideal=\frac{1}{k}\sum\limits_{j=1}^{k}\Tr{\Pi_\abst \Psi_{\bn_j}}$ and $k=4$ (resp. $k=6$) for the maximum likelihood (resp. fidelity) score.

	In our experimental setup, we prepare $N_{\bn_j}=N_0=20000$ photons for each input direction $\bn_j$, and obtain either a guess $\hbn_i$ or an abstention output. Let $\{\{N_{\bn_j,\hbn_i}\},N_{\bn_j,\abst}\}$ be the respective number of time these events happened ($N_{\bn_j}=\sum_i N_{\bn_j,\hbn_i} + N_{\bn_j,\abst}$).
	The experimental score is given by
	\begin{equation}\label{eq: score_exp}
		\begin{aligned}
			s_\ex&=\frac{\sum\limits_{j=1}^{k}\sum\limits_{i=1}^{4}s(\bn_j,\hbn_i) N_{\bn_j,\hbn_i}}{\sum\limits_{j=1}^{k}\sum\limits_{i=1}^{4} N_{\bn_j,\hbn_i}}\\
			&=\frac{\sum\limits_{j=1}^{k}\sum\limits_{i=1}^{4}s(\bn_j,\hbn_i) N_{\bn_j,\hbn_i}}{\sum\limits_{j=1}^{k}{(N_0-N_{\bn_j,\abst})}}\\
			&=\frac{1}{1-\frac{1}{k}\sum\limits_{j=1}^{k}{\frac{N_{\bn_j,\abst}}{N_0}}}\frac{1}{k}\sum\limits_{j=1}^{k}\sum\limits_{i=1}^{4} s(\bn_j,\hbn_i) \frac{N_{\bn_j,\hbn_i}}{N_0}\\
			&=\frac{1}{1-Q_\ex}\frac{1}{k}\sum\limits_{j=1}^{k}\sum\limits_{i=1}^{4} s(\bn_j,\hbn_i) \hp(\bn_j,\hbn_i),
		\end{aligned}
	\end{equation}
	where $\hp(\bn_j,\hbn_i)=\frac{N_{\bn_j,\hbn_i}}{N_0}$ is the  frequency of guess $\hbn_i$ when the input direction is $\bn_j$, 
	and $Q_\ex=\frac{1}{k}\sum\limits_{j=1}^{k}{\frac{N_{\bn_j,\abst}}{N_0}}=\frac{1}{k}\sum\limits_{j=1}^{k} \hp(\bn_j,\abst)$ is the experimental abstention rate, where $\hp(\bn_j,\abst)$ is the abstention frequency for $\bn_j$.
	
\begin{figure}[htbp]
		\center{\includegraphics[width=\linewidth]{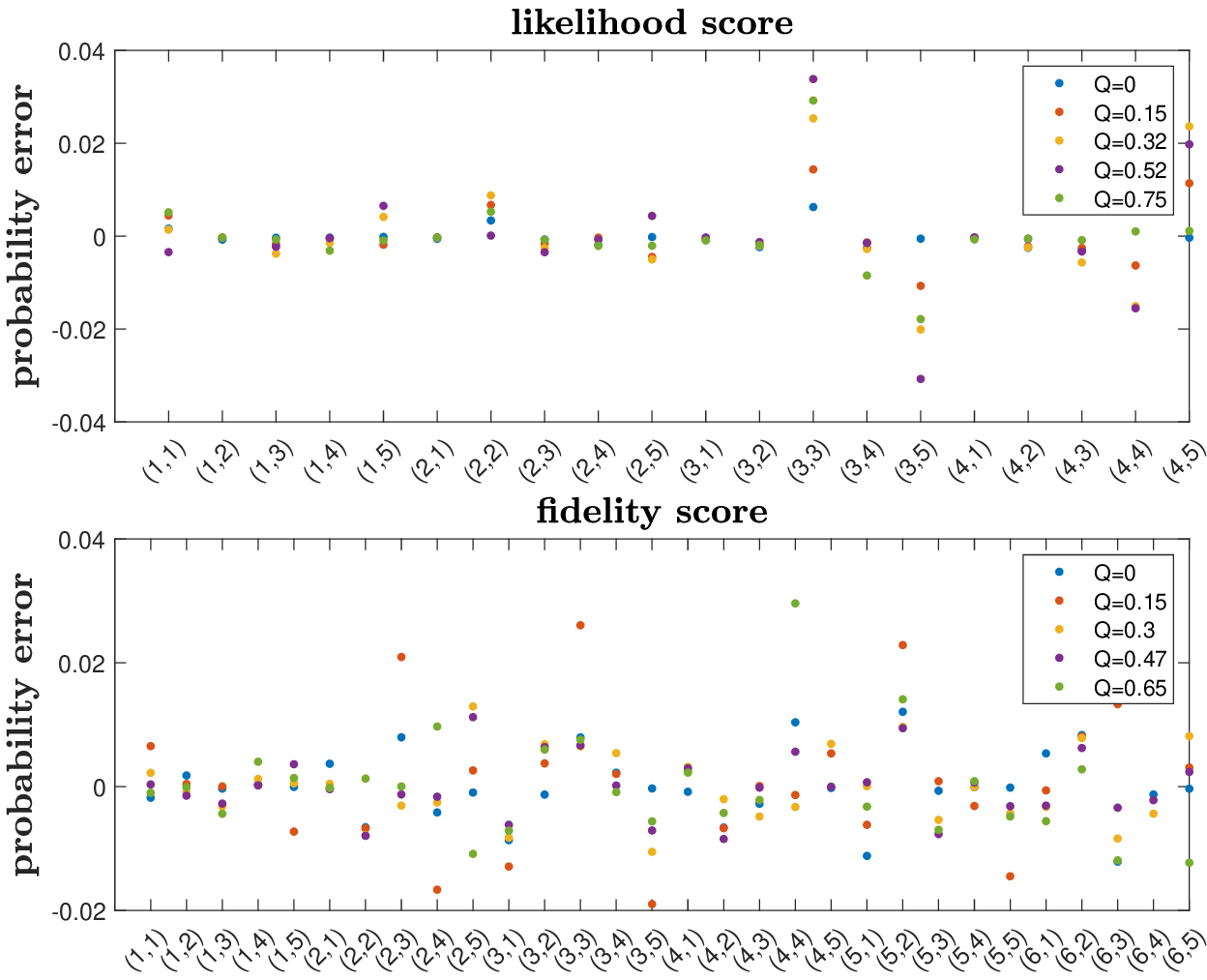}}
		\caption{\label{fig: ProbabilityError} Experimental errors of the probabilities $\Tr{\Pi_i \Psi_{\bn_j}}$, where $\Pi_i=\Pi_{\hbn_i}$ for i=1,...,4 and $\Pi_5=\Pi_\abst$. The coordinates of x-axis represent all the combinations $(\Psi_{\bn_j}, \Pi_i)$. Experimental errors for different abstention rates are marked with different colors, which shows approximate independence with $Q$}.
	\end{figure}

	To analyze the deviation, we introduce $r_{\ideal}=\frac{1}{k}\sum\limits_{j=1}^{k}\sum\limits_{i=1}^{4} s(\bn_j,\hbn_i) \Tr{\Pi_{\hbn_i} \Psi_{\bn_j}}$ and $r_{\ex}=\frac{1}{k}\sum\limits_{j=1}^{k}\sum\limits_{i=1}^{4} s(\bn_j,\hbn_i) \hp(\bn_j, \hbn_i)$. Then we have
	\begin{equation}
		\begin{aligned}
			\Delta s&=s_\ex-s_\ideal \\
			&= \frac{1}{1-Q_\ex} r_{\ex}-\frac{1}{1-Q_\ideal}r_{\ideal}\\
			&=\frac{1}{1-Q_\ex}(r_{\ex}-r_{\ideal})+\frac{Q_\ex-Q_\ideal}{(1-Q_\ex)(1-Q_\ideal)}r_{\ideal}\\
			&=\frac{1}{1-Q_\ex}(\Delta r+ s_\ideal \Delta Q),
		\end{aligned}
	\end{equation}       
	where $s_\ideal$ is independent of $Q$, $\Delta r=r_{\ex}-r_{\ideal}=\frac{1}{k}\sum\limits_{j=1}^{k}\sum\limits_{i=1}^{4} s(\bn_j,\hbn_i) [\hp(\bn_j, \hbn_i)-\Tr{\Pi_{\hbn_i} \Psi_{\bn_j}}]$ and $\Delta Q=Q_\ex-Q_\ideal=\frac{1}{k}\sum\limits_{j=1}^{k} [\hp(\bn_j,\abst)-\Tr{\Pi_\abst \Psi_{\bn_j}}]$.
	$\Delta r, \Delta Q$ depend only on the experimental error of the probabilities $\Tr{\Pi_{\hbn_i} \Psi_{\bn_j}}$ and $\Tr{\Pi_\abst \Psi_{\bn_j}}$, which come from the  imperfection of the experimental apparatus and are with good approximation independent from $Q$, as shown in \fref{fig: ProbabilityError}. However, the term $\frac{1}{1-Q_\ex}$ amplifies this experimental error, thus $|\Delta s|$ is inclinded to increase with the growth of $Q_\ex$.
	
\end{document}